\documentclass[aps,prl,twocolumn,superscriptaddress,showpacs]{revtex4}
\usepackage[english]{babel}
\usepackage{ucs}
\usepackage[utf8x]{inputenc}
\usepackage{amsmath}
\usepackage{amsfonts}
\usepackage{amssymb}
\usepackage{graphicx}
\usepackage{bm}
\usepackage{bbm}
\usepackage{empheq}
\usepackage{color}

\definecolor{jrp}{rgb}{.05,.80,.8}

\definecolor{mjg}{rgb}{.08,.05,.8}

\definecolor{yyl}{rgb}{.8,.05,.08}

\newcommand{\delete}[1]{{}}

\begin{document}

\title{Threshold Dynamics of a Semiconductor Single Atom Maser}

\author{Y.-Y. Liu}
\author{J. Stehlik}
\author{C. Eichler}
\author{X. Mi}
\author{T. Hartke}
\affiliation{Department of Physics, Princeton University, Princeton, New Jersey 08544, USA}
\author{M.~J.~Gullans}
\author{J. M. Taylor}
\affiliation{Joint Quantum Institute and Joint Center for Quantum Information and Computer Science, NIST and University of Maryland, College Park, Maryland 20742, USA}
\author{J. R. Petta}
\affiliation{Department of Physics, Princeton University, Princeton, New Jersey 08544, USA}

\date{\today}

\begin{abstract}
We demonstrate a single-atom maser consisting of a semiconductor double quantum dot (DQD) that is embedded in a high quality factor microwave cavity. A finite bias drives the DQD out of equilibrium, resulting in sequential single electron tunneling and masing. We develop a dynamic tuning protocol that allows us to controllably increase the time-averaged repumping rate of the DQD at a fixed level detuning, and quantitatively study the transition through the masing threshold. We further examine the crossover from incoherent to coherent emission by measuring the photon statistics across the masing transition. The observed threshold behavior is in agreement with an existing single atom maser theory when small corrections from lead emission are taken into account.  
\end{abstract}


\maketitle

Examining photon emission from single quantum emitters provides a window into the interaction between light and matter. The first single atom maser was realized by passing single Rydberg atoms -- which provided a transient gain medium -- through a superconducting cavity \cite{Meschede1985}. 
This cavity quantum electrodynamics (QED) approach provides a template for a variety of single emitter lasing experiments involving optical cavities that are coupled to either natural or artificial atoms \cite{mckeever2003, Reitzenstein2008, Dubin2010, Haroche2013}. Of particular interest is the observation of non-classical optical phenomena, such as Fock state generation and thresholdless lasing \cite{Rice1994, mckeever2003, Reitzenstein2008, Dubin2010, Haroche2013}. In the microwave domain, circuit-QED has enabled dramatic improvements in the coupling between solid-state devices and microwaves, with the demonstration of single photon sources, tomography of itinerant photon states, and even the stabilization of cat states of light \cite{Houck2007, Bozyigit2011, Hofheinz2009, Hofheinz2008, Vlastakis2013}. An on-chip single atom amplifier and maser have also been demonstrated using voltage-biased superconducting junctions \cite{Astafiev2010, Astafiev2007}. 

Semiconductor double quantum dots (DQDs) have been placed in microwave cavities with charge-cavity coupling rates $g_c/2\pi$ = 10 -- 100 MHz \cite{Frey2012, Petersson2012, Toida2013, Viennot2014, Deng2015}, and the strong-coupling regime has recently been achieved in three separate experiments \cite{Mi2017, Stockklauser2017, Bruhat2016}. DQDs allow a great level of experimental control, as their energy level structure is electrically tunable \cite{Wiel2002, Fujisawa1998}. Furthermore, non-equilibrium physics can be explored by applying a source-drain bias across the DQD or by periodically driving the energy level detuning $\epsilon$ \cite{Viennot2014,Liu2015, Stehlik2016}.  These characteristics have enabled a wide range of quantum optics experiments with DQDs, such as photon emission between hybridized single-electron states \cite{Liu2014, Gullans2015, Stockklauser2015}. Masing can be observed in these systems when the gain exceeds the loss, as recently demonstrated by placing two voltage biased InAs DQDs in a microwave cavity \cite{Liu2015}.

In this Letter, we examine the threshold dynamics of a semiconductor single atom maser (SeSAM) consisting of a single DQD that is embedded in a microwave cavity and driven by single electron tunneling events. In contrast with previous experiments that required multiple emitters to exceed the masing threshold \cite{Liu2015}, we demonstrate masing with a single DQD emitter through improvements in the cavity quality factor {$Q_{\rm c}$} and $g_{\rm c}$. We introduce a dynamic tuning protocol that changes the effective repumping rate of the DQD, allowing us to directly observe the transition from below threshold, where the system is dominated by incoherent emission, to above threshold, where the emission is coherent. The threshold behavior is in qualitative agreement with a single atom maser theory from atomic physics \cite{Rice1994}. We obtain quantitative agreement with the data by including photoemission events that originate from tunneling between the DQD and the source-drain electrodes \cite{Childress2004, Jin2011, Kulkarni2014, Gullans2015, Stace2016}. 

\begin{figure}[t]
  \begin{center}
		\includegraphics[width=\columnwidth]{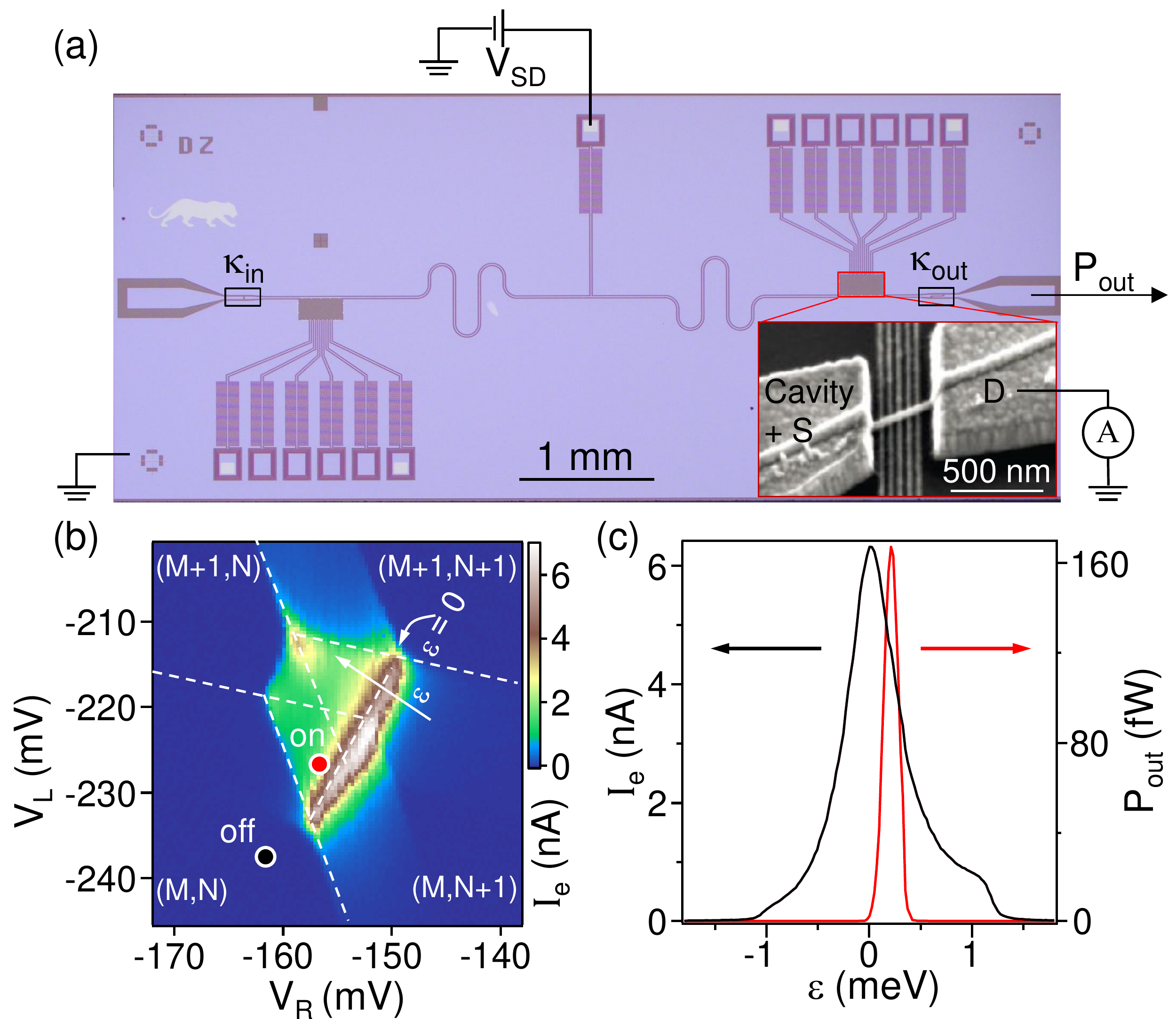}
  \caption{\label{Fig: scheme} (a) Optical micrograph of the DQD SeSAM. The source-drain bias $V_{\rm SD}$ is applied through a $LC$ filtered bias tee. The cavity is connected to input and output ports with coupling rates $\kappa_{\rm in}$ and $\kappa_{\rm out}$. Inset: Scanning electron microscope image of an InAs nanowire DQD. (b) $I_e$ as a function of $V_L$ and $V_R$ with $V_{\rm SD} = 2$ mV. (c) $I_e$ (black) and $P_{\rm out}$ (red) plotted as a function of $\epsilon$. The maximum output power for this device tuning configuration is $P_{\rm out}$ = 160 fW.}
  \end{center}
\end{figure}

The SeSAM consists of a DQD that is embedded in a half-wavelength $\lambda/2$ superconducting cavity [Fig.\ 1(a)]. The cavity has a resonance frequency $f_{\rm c} = 7.5$ GHz and total decay rate $\kappa_{\rm tot}/2\pi = 1.5$ MHz, corresponding to $Q_{\rm c} \approx 5000$. The inset of Fig.\ 1(a) shows a scanning electron microscope image of a single InAs nanowire that is placed on top of 5 metallic gates that are electrically biased to form a double well confinement potential along the length of the nanowire \cite{Nadj-Perge2010}. Charges trapped in the double well have a dipole moment that directly interacts with the cavity field, resulting in a large $g_c/2\pi \approx$ 70 MHz \cite{Petersson2012, Liu2014, Stehlik2016}. In comparison with previous work, we can achieve masing with a single DQD emitter \cite{Nadj-Perge2010, Petersson2012, Liu2014}. The experiment has been improved by reducing microwave leakage through dc gate biasing lines and increasing $g_c$ by fabricating devices with a smaller gap between the source and drain electrodes \cite{Stehlik2016}.

The SeSAM is powered by a source-drain bias $V_{\rm SD} = 2$ mV that is applied to the DQD via the $LC$ filtered bias tee connected to the cavity voltage node \cite{Mi2016APL}. Figure 1(b) shows the resulting current $I_e$ as a function of the gate voltages $V_L$ and $V_R$. Charge states are labeled $(N_L,N_R)$, where $N_{\rm L(R)}$ indicates the electron number in the left (right) dot. Sequential tunneling events are only allowed within the finite bias triangles [as delineated by the dashed lines in Fig.\ 1(b)]. Co-tunneling current is observed outside of the finite bias triangles due to the large tunnel coupling to the leads. As shown in previous work, interdot tunneling leads to photon emission into the cavity mode \cite{Jin2011, Liu2014, Stockklauser2015, Gullans2015}. The resulting cavity field is probed by measuring the power emitted from the cavity output port $P_{\rm out}$ using heterodyne detection (see Ref.\ \cite{Stehlik2016} for measurement details).

Evidence for photon emission due to electron tunneling is shown in Fig.\ 1(c), where $P_{\rm out}$ and $I_{e}$ are plotted as a function of DQD energy level detuning $\epsilon$. The source-drain bias results in a peak current of 6 nA at $\epsilon = 0$, where the left and right dot energy levels are resonant. In contrast, the output power peaks at $\epsilon = 0.3$ meV due to a large phonon sideband, where the current $I_{\rm on}$ = 4 nA and $P_{\rm out} = 160$ fW. For the conditions that result in the maximum output power, the level detuning is approximately 9 times the cavity photon energy (30 $\mu$eV for a 7.5 GHz cavity). Energy is conserved by a process where a phonon and photon are simultaneously emitted \cite{Gullans2015}. The source-drain bias repumps the DQD to the excited state at a rate $|I_e/e|$ and generates the population inversion required for stimulated emission. 

\begin{figure}[t]
  \begin{center}
		\includegraphics[width=\columnwidth]{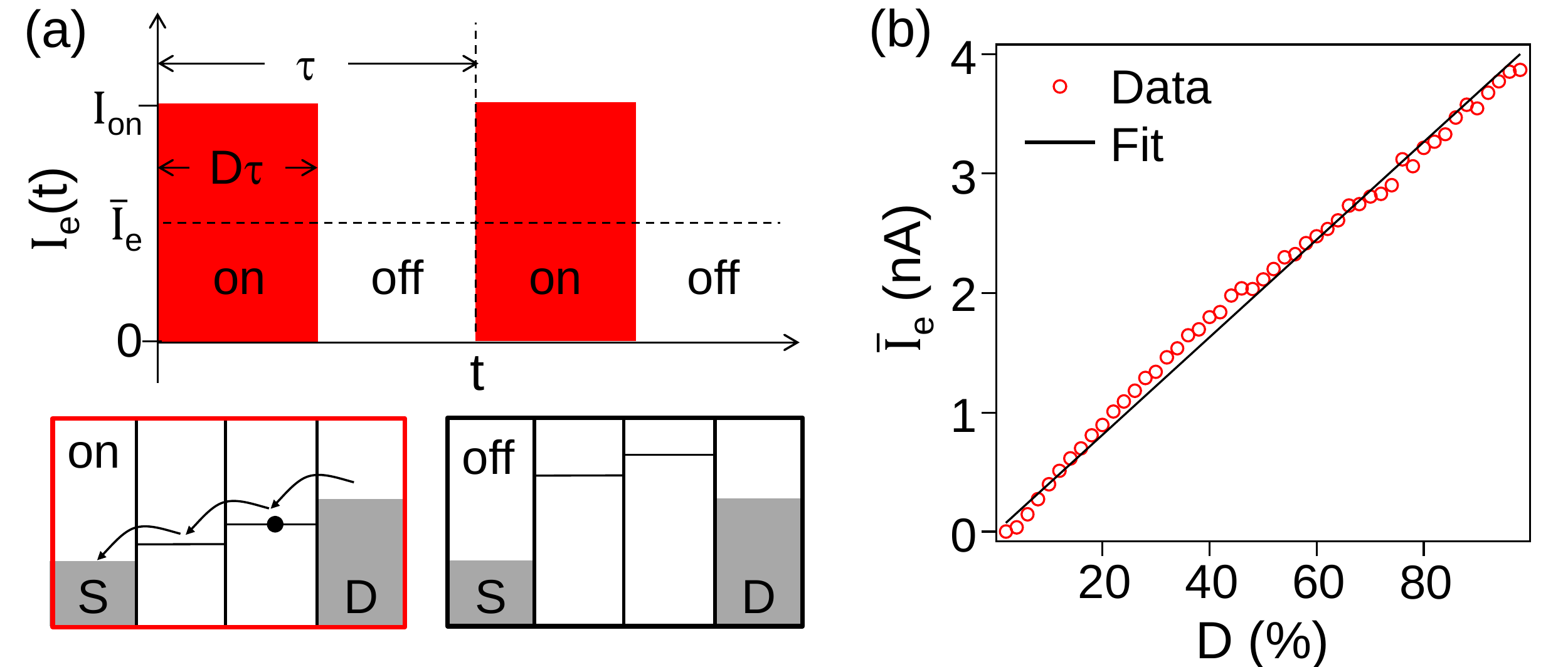}
  \caption{\label{Fig: tuning method} (a) Upper panel: Pulse sequence used to tune the time-averaged current flowing through the device. A square wave with a period $\tau = 50$ ns and duty cycle $D$ is applied to toggle the current on and off. The resulting time-averaged current is $\bar{I}_e = D I_{\rm on}$.  Lower panel: DQD level configuration with the current on (left panel) and off (right panel). (b) The measured $\bar{I}_e$ (red circles) as a function of $D$. The solid line is the prediction $\bar{I}_e= D I_{\rm on}$.
  }
  \end{center}
\end{figure}

The strong emission that is observed is suggestive of above-threshold masing that is triggered by current flow through a single DQD. To investigate the threshold behavior we now measure the statistics of the emitted microwave field as a function of repump rate. In conventional solid state lasers, such as a diode laser, threshold behavior is studied by measuring the emitted power as a function of dc biasing conditions. Such an approach is not directly applicable to DQD devices since the resonant current (and therefore the repump rate) is independent of $V_{\rm SD}$ once the two dots levels are within the transport window \cite{Wiel2002}. Moreover, tuning the tunnel rates also changes the DQD energy level structure and $g_{c}$. We therefore develop a dynamic tuning process that changes the time-averaged current through the DQD. 

As schematically illustrated in Fig.\ \ref{Fig: tuning method}(a), a square wave with a period $\tau = 50$\ ns and duty cycle $D$ is applied to both the left and right gates to toggle the electron current on and off while keeping $\epsilon$ fixed. The locations of the ``on" and ``off" states in the stability diagram are indicated by the red and black dots in Fig.\ \ref{Fig: scheme}(b) and the corresponding energy level configurations are shown in the lower insets of Fig.\ \ref{Fig: tuning method}(a). The time-averaged current through the DQD, $\bar{I}_e$, is plotted as a function of $D$ in Fig.\ 2(b). As expected, the measured current scales linearly with $D$. 

The dynamic tuning process is effective at setting the repump rate because $\kappa_{\rm tot}/2\pi = f_c/Q_c = 1.5$ MHz is much less than the tunneling rate through the DQD, $I_{\rm on}/|e| \approx 25$ GHz. Due to the fact that photon emission is driven by single electron tunneling events, the effective repump rate is proportional to the average current when $|e|/|I_{\rm on}| \ll \tau< 1/\kappa_{\rm tot}$ \cite{SOM}. 

\begin{figure*}[t]
	\begin{center}
		\includegraphics[width=2\columnwidth]{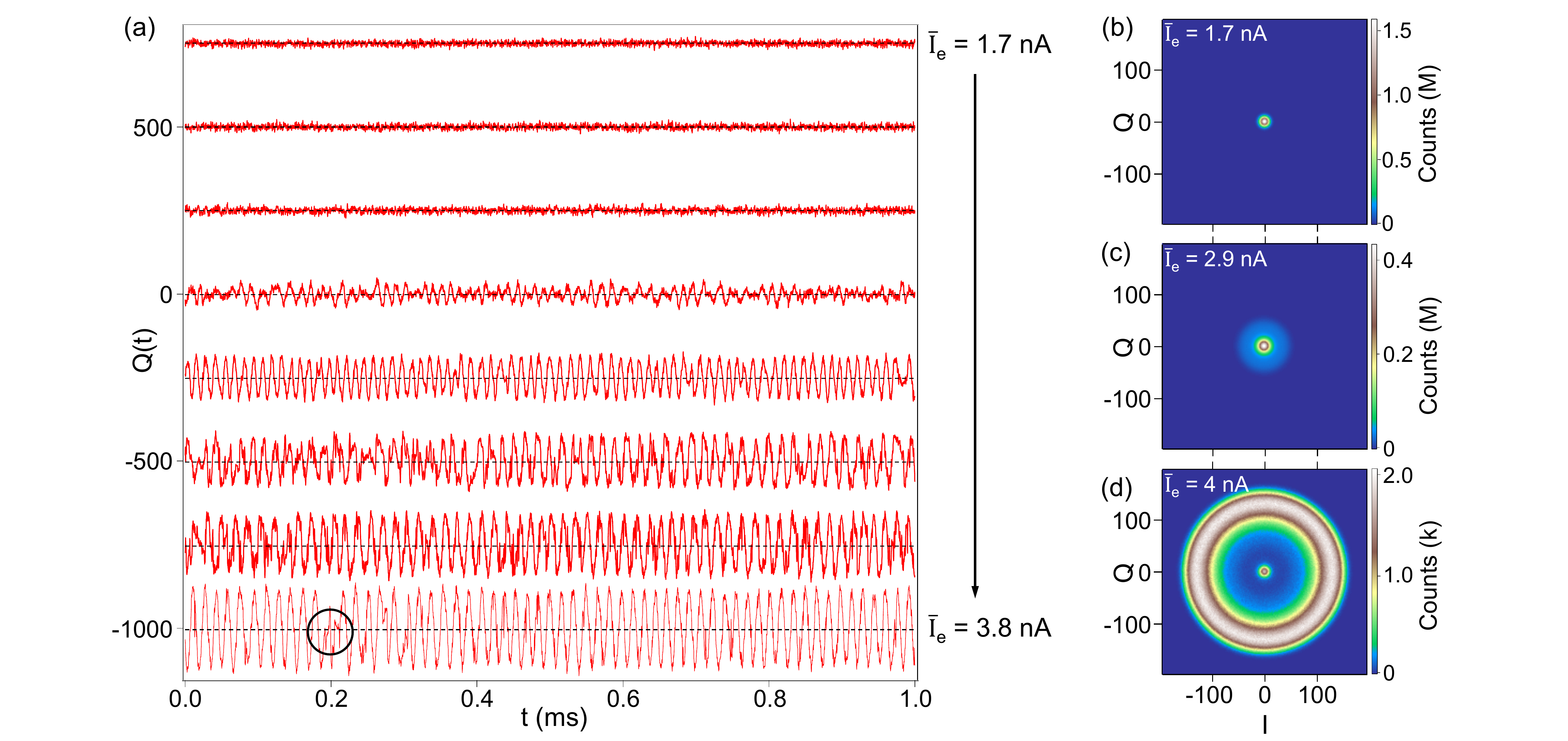}
		\caption{\label{Fig: time traces} (a) Time series $Q(t)$ with $\bar{I}_e$ ranging from 1.7\ nA to 3.8\ nA with a step size of 0.3 nA. The data are offset along the x-axis by 250 in $Q$ for clarity. The emission amplitude increases with increasing $\bar{I}_e$. 
		(b--d) $IQ$ histograms acquired with $\bar{I}_e$ = 1.7, 2.9, and 4 nA. (b) The emission is centered around ($I$, $Q$) = (0,0) at small currents. (c) The distribution gradually broadens as $\bar{I}_e$ is increased. (d) With $\bar{I}_e$ = 4 nA, the histogram has a qualitatively different ring shape that is indicative of above threshold maser emission.}
	\end{center}	
\end{figure*}

The time-series of the demodulated quadrature-phase component of the cavity field $Q(t)$ qualitatively illustrates the crossover from below threshold to above threshold behavior as $\bar{I}_e$ is increased [Fig.\ 3(a)].  When $\bar{I}_e<2.5$ nA the output is thermal noise, which is mainly attributed to background noise in the amplification chain. When $\bar{I}_e > 2.5$ nA periodic voltage oscillations become visible, with an amplitude that increases with $\bar{I}_e$. The oscillating field is indicative of coherent emission. We note that the maser occasionally blinks off even for large $\bar{I}_e$ (see for example $t = 0.2$ ms at $\bar{I}_e = 3.8$ nA). We attribute the blinking to large charge fluctuations that shift the DQD level detuning \cite{Liu2015}. Similar behavior has been observed in other solid state lasers \cite{Siegman1986}. These data show that the dynamical detuning method effectively changes the DQD repump rate.

The maser emission statistics can be quantitatively studied by measuring histograms of the output field as a function of increasing $\bar{I}_e$. For each value of $\bar{I}_e$, we sample the in-phase and quadrature-phase components of the cavity field ($I$ and $Q$) at a rate of 12.3 MHz and then plot $1.7\times10^7$ samples in a two-dimensional histogram \cite{Liu2015, Stehlik2016}.  Histograms with $\bar{I}_e=1.7$, 2.9 and 4 nA are plotted in Figs.\ \ref{Fig: time traces}(b--d). With a small current of $\bar{I}_e = 1.7$ nA, the histogram is centered within a narrow range of the origin in the $IQ$ plane, as the detected field is dominated by the detection background noise. With increasing current, $\bar{I}_e=2.9$ nA, the histogram broadens out into a larger range as shown in Fig.\ \ref{Fig: time traces}(c). For $\bar{I}_e = 4$ nA ($D=100\%$), the $IQ$ histogram has a ring shape that is indicative of above-threshold maser emission [Fig.\ \ref{Fig: time traces}(d)] and a small thermal component around ($I$, $Q$) = (0,0). As noted in previous work, the events around $(I,Q) = (0,0)$ in Fig.\ \ref{Fig: time traces}(d) are attributed to blinking events (circled in black) that are visible in the time series data of Fig.\ \ref{Fig: time traces}(a) \cite{Liu2015}.

\begin{figure}[t]
	\begin{center}
		\includegraphics[width=\columnwidth]{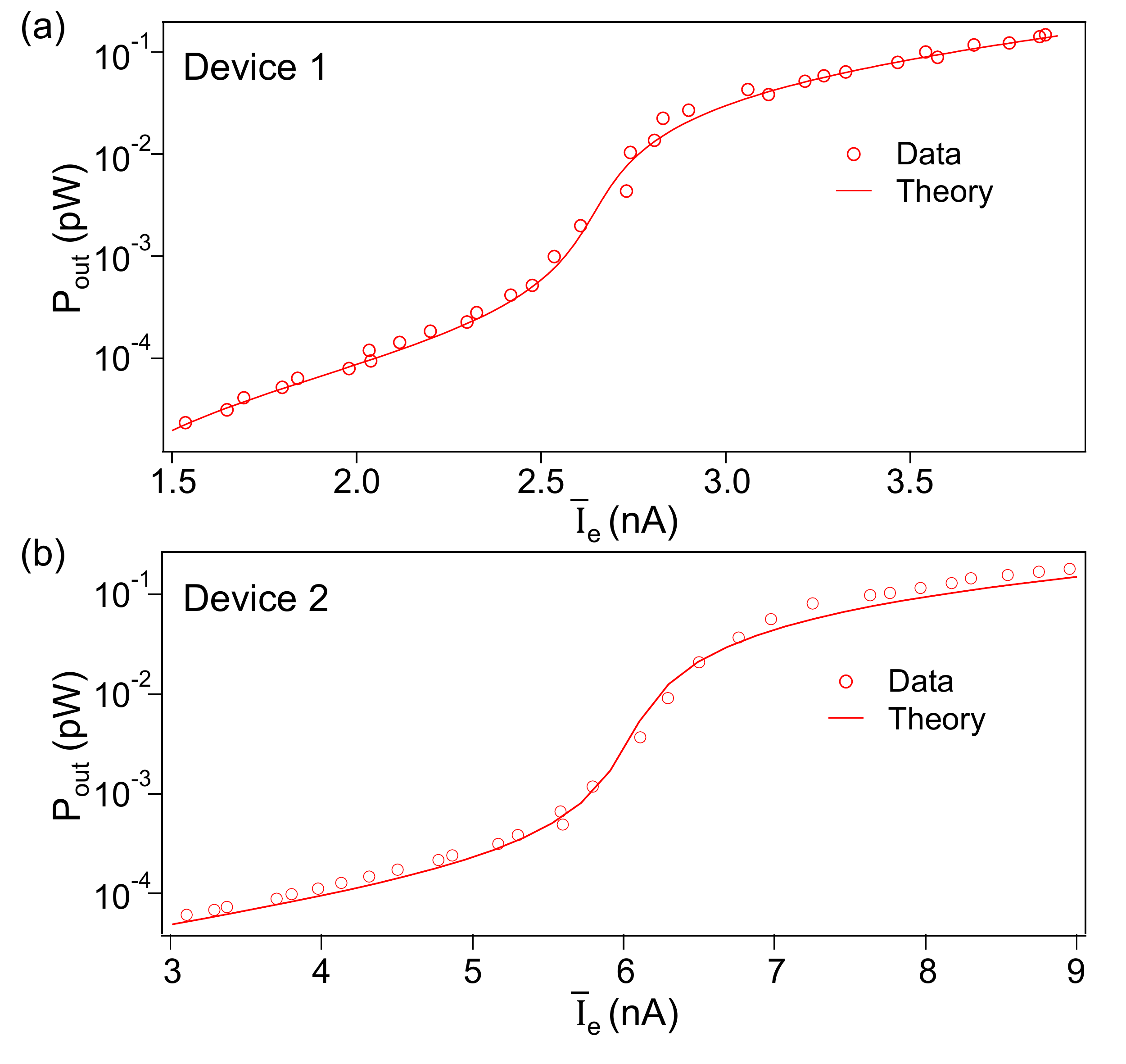}
		\caption{\label{Fig: threshold} $P_{\rm out}$ (red circles) as a function of $\bar{I}_e$ and fits to theory (solid lines) for two different devices. (a) Device 1 crosses threshold around $\bar{I}_e \approx 2.6$ nA. (b) Device 2 crosses threshold around $\bar{I}_e \approx 6.1$ nA. Both devices exhibit qualitatively similar threshold behavior.}
	\end{center}	
\end{figure}

Measurements of $P_{\rm out}$ also provide insight into the threshold behavior and can be compared with existing single atom maser theories \cite{Rice1994}. Figure \ref{Fig: threshold} plots $P_{\rm out}$ as a function of $\bar{I}_e$ for two different devices. Focusing on Fig.\ \ref{Fig: threshold}(a), for small $\bar{I}_e= 1.5$ nA we measure $P_{\rm out}\approx 10^{-5}$ pW. $P_{\rm out}$ gradually increases with increasing $\bar{I}_e$ until $\bar{I}_e$ $\approx$ 2.5 nA. There is a dramatic factor of $\approx$ 50 increase in $P_{\rm out}$ in the range $2.5 < \bar{I}_e < 2.8$ nA. For $\bar{I}_e> 2.8$ nA we find that $P_{\rm out}$ increases slowly again with  $\bar{I}_e$. These data indicate that the maser crosses threshold when $\bar{I}_e \approx 2.5\sim 2.8$ nA and is well above threshold for $\bar{I}_e > 2.8$ nA.

We can understand the threshold behavior with an expanded form of the semiclassical theory of a single-atom laser  \cite{Rice1994}. In particular, we model the DQD SeSAM using semiclassical laser equations that account for photon emission events during interdot tunneling, as well as during tunneling onto and off of the leads
\begin{align}
\dot{u} &= - [\gamma+ \Gamma_p(n_c) + R] u + \Gamma_p(n_c) - R n_c u,\\
\dot{n}_c &= - \kappa_{\rm tot} n_c + R n_c u + R u, \\
\Gamma_p&(n_c)=[1+\alpha(n_c+1)] \Gamma_p^0,
\end{align}
where $u$ is the population of the state with one electron in the right dot (see Fig. 2), $\Gamma_p(n_c)$ is the tunneling rate between the source-drain electrodes and the DQD states, $\Gamma_p^0$ is the  bare tunneling rate, $\alpha$ is the fraction of lead tunneling events that result in photon emission, $\gamma$ is the decay rate of the upper state of the DQD into modes (e.g., phonons) other than the cavity, $R$ is the photon emission rate from the upper state of the DQD, and $\kappa_{\rm tot}$ is the total cavity decay rate.  The dependence of the lead tunneling on cavity photon number arises from photon-assisted tunneling events between the leads and the DQD.  When $\alpha=0$, Eqs.~(1--3) reduce to the conventional semiclassical theory of the single-atom laser \cite{Rice1994}. Spontaneous emission is accounted for by the terms where $R$ and $\alpha$ appear independently of $n_c$.  

We model the tuning cycle by coarse graining Eqs.~(1--2) over one period with a duty cycle $D$, which effectively replaces $u$ in Eq.~(2) by $D u$.  Similarly, we model the average current through the DQD as $\bar{I}_e = e D (\gamma +R+ R n_c) u$.  
The crucial parameter that relates $\bar{I}_e$ to $P_{\rm out}$ is the fraction of electron tunneling events through the DQD that result in photon emission, which we define as $\beta$.  We can estimate  $\beta$ by noting that the threshold current in our model is given by $I_{\textrm{th}} = |e| \kappa_{\textrm{tot}} / \beta$ with  $\beta = R/\gamma + 2 \alpha$.  The full expression for $P_{\rm out}$ is derived in the supplemental material \cite{SOM}, with the final expression given here:
\begin{equation}
\begin{split}
P_{\textrm{out}} &= \frac{G^T_{\rm out}}{G^E_{\rm out}} \frac{h f_c \kappa_{\textrm{out}}}
{2 \xi R/\gamma} 
\left[     \frac{\bar{I}_e}{I_{\textrm{th}}} - 1 + \sqrt{\left(\frac{\bar{I}_e}{I_{\textrm{th}}} - 1 \right)^2 + \frac{4 \xi \bar{I}_e}{I_{\textrm{th}}}}\right].
\end{split}
\end{equation}
\noindent
$\kappa_{\rm out}$ is the coupling rate to the output port of the cavity and is designed to be $\kappa_{\rm out}/2\pi = 0.8$ MHz for both devices.  Here we have also defined a correction parameter associated with the lead emission $\xi = 1 - \frac{2 \alpha \bar{I}_e}{|e| \kappa_{\textrm{tot}}}$. The prefactor ${G^T_{\rm out}}/{G^E_{\rm out}}$ accounts for the systematic error in the total gain of the detection chain, $G_{\rm out}$, from the output port of the cavity to the top of the fridge. 

\begin{table}
	\begin{center}
		\caption{Fit parameters for the threshold behavior in Fig.~\ref{Fig: threshold}}
		\begin{ruledtabular}
			\begin{tabular}{l l  l  l }
				 &  & Device 1 &  Device 2 \\\hline 
				Measured & $f_{\rm c}$ & 7.5 GHz & 7.6 GHz \\
				parameters& $\kappa_{\rm tot}/2\pi$ & 1.5 MHz & 1.8 MHz \\\hline 
				Calibrated  & $G^E_{\rm out}$  & 74.5 dB & 72.8 dB \\\hline 
				Free  & $G^T_{\rm out}$  & 76.3 dB& 76.2 dB\\
				parameters& $I_{\rm th}$ & 2.63  nA & 6.14  nA  \\
				& $\alpha$ & 1.2$\times10^{-4}$ &  0.3$\times10^{-4}$\\\hline 
				Calculated & $\beta = \kappa_{\rm tot}/|e|I_{\rm th}$   & 5.7$\times10^{-4}$ & 2.9$\times10^{-4}$\\
				 & $R/\gamma = \beta- 2\alpha$  & 3.3$\times10^{-4}$ &  2.3$\times10^{-4}$ 
				\label{Table: threshold fits}
			\end{tabular}
		\end{ruledtabular}
	\end{center}
\end{table}
	
The theoretical prediction is in good agreement with data from two devices, as shown in Fig.~4. Fit parameters are listed in Table\ \ref{Table: threshold fits}. From calibrations of the amplifier gain, and room temperature measurements of the losses in the coax lines, we estimate the total gain of the amplification chain to be $G^E_{\rm out} = 74.5$~dB for device 1, and $G^E_{\rm out} = 72.8$~dB for device 2 in another cool down. Given experimental uncertainties in $\kappa_{\rm out}$, losses in the device packaging, and the temperature dependent losses in the coax lines, these values are in overall agreement with the best fit value $G^T_{\rm out} = 76.3$~dB for device 1 and $G^T_{\rm out} = 76.2$~dB for device 2 \cite{LiuPRA2015, Stehlik2016}.  The quantitative agreement of this model with the data suggests that lead emission events play an important role in the charge-cavity dynamics of our device \cite{Childress2004, Jin2011, Kulkarni2014, Gullans2015, Stace2016}. The reader is referred to the supplemental material for a comparison to the standard single atom laser theory, which does not account for lead emission \cite{SOM}.

In conclusion, we have measured the threshold dynamics of a semiconductor single atom maser, which allows for investigations of maser physics in the simplest case of a single emitter in a cavity. Photon emission in the SeSAM is generated by single electron tunneling events. By implementing a dynamic tuning protocol, we quantitatively analyze the behavior of the SeSAM as it crosses threshold. The data are in agreement with a modified single atom maser theory that includes a correction due to lead emission.

\begin{acknowledgments}
Research was supported by the Gordon and Betty Moore Foundation’s EPiQS Initiative through Grant GBMF4535, with partial support from the National Science Foundation (Grants No.\ DMR-1409556 and DMR-1420541). Devices were fabricated in the Princeton University Quantum Device Nanofabrication Laboratory.
\end{acknowledgments}

\bibliographystyle{apsrev_lyy2017}
\bibliography{SAM_locking_v4}

\begin{thebibliography}{37}
\expandafter\ifx\csname natexlab\endcsname\relax\def\natexlab#1{#1}\fi
\expandafter\ifx\csname bibnamefont\endcsname\relax
  \def\bibnamefont#1{#1}\fi
\expandafter\ifx\csname bibfnamefont\endcsname\relax
  \def\bibfnamefont#1{#1}\fi
\expandafter\ifx\csname citenamefont\endcsname\relax
  \def\citenamefont#1{#1}\fi
\expandafter\ifx\csname url\endcsname\relax
  \def\url#1{\texttt{#1}}\fi
\expandafter\ifx\csname urlprefix\endcsname\relax\def\urlprefix{URL }\fi
\providecommand{\bibinfo}[2]{#2}
\providecommand{\eprint}[2][]{\url{#2}}

\bibitem[{\citenamefont{Meschede et~al.}(1985)\citenamefont{Meschede, Walther,
  and M\"uller}}]{Meschede1985}
\bibinfo{author}{\bibfnamefont{D.}~\bibnamefont{Meschede}},
  \bibinfo{author}{\bibfnamefont{H.}~\bibnamefont{Walther}}, \bibnamefont{and}
  \bibinfo{author}{\bibfnamefont{G.}~\bibnamefont{M\"uller}}, One-Atom Maser,
  \bibinfo{journal}{Phys. Rev. Lett.} \textbf{\bibinfo{volume}{54}},
  \bibinfo{pages}{551} (\bibinfo{year}{1985}).

\bibitem[{\citenamefont{McKeever et~al.}(2003)\citenamefont{McKeever, Boca,
  Boozer, Buck, and Kimble}}]{mckeever2003}
\bibinfo{author}{\bibfnamefont{J.}~\bibnamefont{McKeever}},
  \bibinfo{author}{\bibfnamefont{A.}~\bibnamefont{Boca}},
  \bibinfo{author}{\bibfnamefont{A.~D.} \bibnamefont{Boozer}},
  \bibinfo{author}{\bibfnamefont{J.~R.} \bibnamefont{Buck}}, \bibnamefont{and}
  \bibinfo{author}{\bibfnamefont{H.~J.} \bibnamefont{Kimble}}, Experimental
  Realization of a One-Atom Laser in the Regime of Strong Coupling,
  \bibinfo{journal}{Nature (London)} \textbf{\bibinfo{volume}{425}},
  \bibinfo{pages}{268} (\bibinfo{year}{2003}).

\bibitem[{\citenamefont{Reitzenstein et~al.}(2008)\citenamefont{Reitzenstein,
  Heindel, Kistner, Rahimi-Iman, Schneider, H\"{o}fling, and
  Forchel}}]{Reitzenstein2008}
\bibinfo{author}{\bibfnamefont{S.}~\bibnamefont{Reitzenstein}},
  \bibinfo{author}{\bibfnamefont{T.}~\bibnamefont{Heindel}},
  \bibinfo{author}{\bibfnamefont{C.}~\bibnamefont{Kistner}},
  \bibinfo{author}{\bibfnamefont{A.}~\bibnamefont{Rahimi-Iman}},
  \bibinfo{author}{\bibfnamefont{C.}~\bibnamefont{Schneider}},
  \bibinfo{author}{\bibfnamefont{S.}~\bibnamefont{H\"{o}fling}},
  \bibnamefont{and} \bibinfo{author}{\bibfnamefont{A.}~\bibnamefont{Forchel}},
  Low Threshold Electrically Pumped Quantum Dot-Micropillar Lasers,
  \bibinfo{journal}{App. Phys. Lett.} \textbf{\bibinfo{volume}{93}},
  \bibinfo{eid}{061104} (\bibinfo{year}{2008}).

\bibitem[{\citenamefont{Dubin et~al.}(2010)\citenamefont{Dubin, Russo, Barros,
  Stute, Becher, Schmidt, and Blatt}}]{Dubin2010}
\bibinfo{author}{\bibfnamefont{F.}~\bibnamefont{Dubin}},
  \bibinfo{author}{\bibfnamefont{C.}~\bibnamefont{Russo}},
  \bibinfo{author}{\bibfnamefont{H.~G.} \bibnamefont{Barros}},
  \bibinfo{author}{\bibfnamefont{A.}~\bibnamefont{Stute}},
  \bibinfo{author}{\bibfnamefont{C.}~\bibnamefont{Becher}},
  \bibinfo{author}{\bibfnamefont{P.~O.} \bibnamefont{Schmidt}},
  \bibnamefont{and} \bibinfo{author}{\bibfnamefont{R.}~\bibnamefont{Blatt}},
  Quantum to Classical Transition in a Single-Ion Laser, \bibinfo{journal}{Nat.
  Phys.} \textbf{\bibinfo{volume}{6}}, \bibinfo{pages}{350}
  (\bibinfo{year}{2010}).

\bibitem[{\citenamefont{Haroche}(2013)}]{Haroche2013}
\bibinfo{author}{\bibfnamefont{S.}~\bibnamefont{Haroche}}, Nobel Lecture:
  Controlling Photons in a Box and Exploring the Quantum to Classical Boundary,
  \bibinfo{journal}{Rev. Mod. Phys.} \textbf{\bibinfo{volume}{85}},
  \bibinfo{pages}{1083} (\bibinfo{year}{2013}).

\bibitem[{\citenamefont{Rice and Carmichael}(1994)}]{Rice1994}
\bibinfo{author}{\bibfnamefont{P.~R.} \bibnamefont{Rice}} \bibnamefont{and}
  \bibinfo{author}{\bibfnamefont{H.~J.} \bibnamefont{Carmichael}}, Photon
  Statistics of a Cavity-QED Laser: A Comment on the Laser-Phase-Transition
  Analogy, \bibinfo{journal}{Phys. Rev. A} \textbf{\bibinfo{volume}{50}},
  \bibinfo{pages}{4318} (\bibinfo{year}{1994}).

\bibitem[{\citenamefont{Houck et~al.}(2007)\citenamefont{Houck, Schuster,
  Gambetta, Schreier, Johnson, Chow, Frunzio, Majer, Devoret, Girvin
  et~al.}}]{Houck2007}
\bibinfo{author}{\bibfnamefont{A.~A.} \bibnamefont{Houck}}, \emph{et~al.},
  Generating Single Microwave Photons in a Circuit, \bibinfo{journal}{Nature
  (London)} \textbf{\bibinfo{volume}{449}}, \bibinfo{pages}{328}
  (\bibinfo{year}{2007}).

\bibitem[{\citenamefont{Bozyigit et~al.}(2011)\citenamefont{Bozyigit, Lang,
  Steffen, Fink, Eichler, Baur, Bianchetti, Leek, Filipp, da~Silva
  et~al.}}]{Bozyigit2011}
\bibinfo{author}{\bibfnamefont{D.}~\bibnamefont{Bozyigit}}, \emph{et~al.},
  Antibunching of Microwave-Frequency Photons Observed in Correlation
  Measurements Using Linear Detectors, \bibinfo{journal}{Nat. Phys.}
  \textbf{\bibinfo{volume}{7}}, \bibinfo{pages}{154} (\bibinfo{year}{2011}).

\bibitem[{\citenamefont{Hofheinz et~al.}(2009)\citenamefont{Hofheinz, Wang,
  Ansmann, Bialczak, Lucero, Neeley, O'Connell, Sank, Wenner, Martinis
  et~al.}}]{Hofheinz2009}
\bibinfo{author}{\bibfnamefont{M.}~\bibnamefont{Hofheinz}}, \emph{et~al.},
  Synthesizing Arbitrary Quantum States in a Superconducting Resonator,
  \bibinfo{journal}{Nature (London)} \textbf{\bibinfo{volume}{459}},
  \bibinfo{pages}{546} (\bibinfo{year}{2009}).

\bibitem[{\citenamefont{Hofheinz et~al.}(2008)\citenamefont{Hofheinz, Weig,
  Ansmann, Bialczak, Lucero, Neeley, O'Connell, Wang, Martinis, and
  Cleland}}]{Hofheinz2008}
\bibinfo{author}{\bibfnamefont{M.}~\bibnamefont{Hofheinz}},
  \bibinfo{author}{\bibfnamefont{E.~M.} \bibnamefont{Weig}},
  \bibinfo{author}{\bibfnamefont{M.}~\bibnamefont{Ansmann}},
  \bibinfo{author}{\bibfnamefont{R.~C.} \bibnamefont{Bialczak}},
  \bibinfo{author}{\bibfnamefont{E.}~\bibnamefont{Lucero}},
  \bibinfo{author}{\bibfnamefont{M.}~\bibnamefont{Neeley}},
  \bibinfo{author}{\bibfnamefont{A.~D.} \bibnamefont{O'Connell}},
  \bibinfo{author}{\bibfnamefont{H.}~\bibnamefont{Wang}},
  \bibinfo{author}{\bibfnamefont{J.~M.} \bibnamefont{Martinis}},
  \bibnamefont{and} \bibinfo{author}{\bibfnamefont{A.~N.}
  \bibnamefont{Cleland}}, Generation of {Fock} States in a Superconducting
  Quantum Circuit, \bibinfo{journal}{Nature (London)}
  \textbf{\bibinfo{volume}{454}}, \bibinfo{pages}{310} (\bibinfo{year}{2008}).

\bibitem[{\citenamefont{Vlastakis et~al.}(2013)\citenamefont{Vlastakis,
  Kirchmair, Leghtas, Nigg, Frunzio, Girvin, Mirrahimi, Devoret, and
  Schoelkopf}}]{Vlastakis2013}
\bibinfo{author}{\bibfnamefont{B.}~\bibnamefont{Vlastakis}},
  \bibinfo{author}{\bibfnamefont{G.}~\bibnamefont{Kirchmair}},
  \bibinfo{author}{\bibfnamefont{Z.}~\bibnamefont{Leghtas}},
  \bibinfo{author}{\bibfnamefont{S.~E.} \bibnamefont{Nigg}},
  \bibinfo{author}{\bibfnamefont{L.}~\bibnamefont{Frunzio}},
  \bibinfo{author}{\bibfnamefont{S.~M.} \bibnamefont{Girvin}},
  \bibinfo{author}{\bibfnamefont{M.}~\bibnamefont{Mirrahimi}},
  \bibinfo{author}{\bibfnamefont{M.~H.} \bibnamefont{Devoret}},
  \bibnamefont{and} \bibinfo{author}{\bibfnamefont{R.~J.}
  \bibnamefont{Schoelkopf}}, Deterministically Encoding Quantum Information
  Using 100-Photon Schr\"{o}dinger Cat States, \bibinfo{journal}{Science}
  \textbf{\bibinfo{volume}{342}}, \bibinfo{pages}{607} (\bibinfo{year}{2013}).

\bibitem[{\citenamefont{Astafiev et~al.}(2010)\citenamefont{Astafiev,
  Abdumalikov, Zagoskin, Pashkin, Nakamura, and Tsai}}]{Astafiev2010}
\bibinfo{author}{\bibfnamefont{O.~V.} \bibnamefont{Astafiev}},
  \bibinfo{author}{\bibfnamefont{A.~A.} \bibnamefont{Abdumalikov}},
  \bibinfo{author}{\bibfnamefont{A.~M.} \bibnamefont{Zagoskin}},
  \bibinfo{author}{\bibfnamefont{Y.~A.} \bibnamefont{Pashkin}},
  \bibinfo{author}{\bibfnamefont{Y.}~\bibnamefont{Nakamura}}, \bibnamefont{and}
  \bibinfo{author}{\bibfnamefont{J.~S.} \bibnamefont{Tsai}}, Ultimate On-Chip
  Quantum Amplifier, \bibinfo{journal}{Phys. Rev. Lett.}
  \textbf{\bibinfo{volume}{104}}, \bibinfo{pages}{183603}
  (\bibinfo{year}{2010}).

\bibitem[{\citenamefont{Astafiev et~al.}(2007)\citenamefont{Astafiev, Inomata,
  Niskanen, Yamamoto, Pashkin, Nakamura, and Tsai}}]{Astafiev2007}
\bibinfo{author}{\bibfnamefont{O.}~\bibnamefont{Astafiev}},
  \bibinfo{author}{\bibfnamefont{K.}~\bibnamefont{Inomata}},
  \bibinfo{author}{\bibfnamefont{A.~O.} \bibnamefont{Niskanen}},
  \bibinfo{author}{\bibfnamefont{T.}~\bibnamefont{Yamamoto}},
  \bibinfo{author}{\bibfnamefont{Y.~A.} \bibnamefont{Pashkin}},
  \bibinfo{author}{\bibfnamefont{Y.}~\bibnamefont{Nakamura}}, \bibnamefont{and}
  \bibinfo{author}{\bibfnamefont{J.~S.} \bibnamefont{Tsai}}, Single
  Artificial-Atom Lasing, \bibinfo{journal}{Nature (London)}
  \textbf{\bibinfo{volume}{449}}, \bibinfo{pages}{588} (\bibinfo{year}{2007}).

\bibitem[{\citenamefont{Frey et~al.}(2012)\citenamefont{Frey, Leek, Beck,
  Blais, Ihn, Ensslin, and Wallraff}}]{Frey2012}
\bibinfo{author}{\bibfnamefont{T.}~\bibnamefont{Frey}},
  \bibinfo{author}{\bibfnamefont{P.~J.} \bibnamefont{Leek}},
  \bibinfo{author}{\bibfnamefont{M.}~\bibnamefont{Beck}},
  \bibinfo{author}{\bibfnamefont{A.}~\bibnamefont{Blais}},
  \bibinfo{author}{\bibfnamefont{T.}~\bibnamefont{Ihn}},
  \bibinfo{author}{\bibfnamefont{K.}~\bibnamefont{Ensslin}}, \bibnamefont{and}
  \bibinfo{author}{\bibfnamefont{A.}~\bibnamefont{Wallraff}}, Dipole Coupling
  of a Double Quantum Dot to a Microwave Resonator, \bibinfo{journal}{Phys.
  Rev. Lett.} \textbf{\bibinfo{volume}{108}}, \bibinfo{pages}{046807}
  (\bibinfo{year}{2012}).

\bibitem[{\citenamefont{Petersson et~al.}(2012)\citenamefont{Petersson, McFaul,
  Schroer, Jung, Taylor, Houck, and Petta}}]{Petersson2012}
\bibinfo{author}{\bibfnamefont{K.~D.} \bibnamefont{Petersson}},
  \bibinfo{author}{\bibfnamefont{L.~W.} \bibnamefont{McFaul}},
  \bibinfo{author}{\bibfnamefont{M.~D.} \bibnamefont{Schroer}},
  \bibinfo{author}{\bibfnamefont{M.}~\bibnamefont{Jung}},
  \bibinfo{author}{\bibfnamefont{J.~M.} \bibnamefont{Taylor}},
  \bibinfo{author}{\bibfnamefont{A.~A.} \bibnamefont{Houck}}, \bibnamefont{and}
  \bibinfo{author}{\bibfnamefont{J.~R.} \bibnamefont{Petta}}, Circuit Quantum
  Electrodynamics with a Spin Qubit, \bibinfo{journal}{Nature (London)}
  \textbf{\bibinfo{volume}{490}}, \bibinfo{pages}{380} (\bibinfo{year}{2012}).

\bibitem[{\citenamefont{Toida et~al.}(2013)\citenamefont{Toida, Nakajima, and
  Komiyama}}]{Toida2013}
\bibinfo{author}{\bibfnamefont{H.}~\bibnamefont{Toida}},
  \bibinfo{author}{\bibfnamefont{T.}~\bibnamefont{Nakajima}}, \bibnamefont{and}
  \bibinfo{author}{\bibfnamefont{S.}~\bibnamefont{Komiyama}}, Vacuum Rabi
  Splitting in a Semiconductor Circuit {QED} System, \bibinfo{journal}{Phys.
  Rev. Lett.} \textbf{\bibinfo{volume}{110}}, \bibinfo{pages}{066802}
  (\bibinfo{year}{2013}).

\bibitem[{\citenamefont{Viennot et~al.}(2014)\citenamefont{Viennot, Delbecq,
  Dartiailh, Cottet, and Kontos}}]{Viennot2014}
\bibinfo{author}{\bibfnamefont{J.~J.} \bibnamefont{Viennot}},
  \bibinfo{author}{\bibfnamefont{M.~R.} \bibnamefont{Delbecq}},
  \bibinfo{author}{\bibfnamefont{M.~C.} \bibnamefont{Dartiailh}},
  \bibinfo{author}{\bibfnamefont{A.}~\bibnamefont{Cottet}}, \bibnamefont{and}
  \bibinfo{author}{\bibfnamefont{T.}~\bibnamefont{Kontos}}, Out-of-Equilibrium
  Charge Dynamics in a Hybrid Circuit Quantum Electrodynamics Architecture,
  \bibinfo{journal}{Phys. Rev. B} \textbf{\bibinfo{volume}{89}},
  \bibinfo{pages}{165404} (\bibinfo{year}{2014}).

\bibitem[{\citenamefont{Deng et~al.}(2015)\citenamefont{Deng, Wei, Johansson,
  Zhang, Li, Li, Cao, Xiao, Tu, Guo et~al.}}]{Deng2015}
\bibinfo{author}{\bibfnamefont{G.-W.} \bibnamefont{Deng}}, \emph{et~al.},
  Charge Number Dependence of the Dephasing Rates of a Graphene Double Quantum
  Dot in a Circuit QED Architecture, \bibinfo{journal}{Phys. Rev. Lett.}
  \textbf{\bibinfo{volume}{115}}, \bibinfo{pages}{126804}
  (\bibinfo{year}{2015}).

\bibitem[{\citenamefont{Mi et~al.}(2017{\natexlab{a}})\citenamefont{Mi, Cady,
  Zajac, Deelman, and Petta}}]{Mi2017}
\bibinfo{author}{\bibfnamefont{X.}~\bibnamefont{Mi}},
  \bibinfo{author}{\bibfnamefont{J.~V.} \bibnamefont{Cady}},
  \bibinfo{author}{\bibfnamefont{D.~M.} \bibnamefont{Zajac}},
  \bibinfo{author}{\bibfnamefont{P.~W.} \bibnamefont{Deelman}},
  \bibnamefont{and} \bibinfo{author}{\bibfnamefont{J.~R.} \bibnamefont{Petta}},
  Strong Coupling of a Single Electron in Silicon to a Microwave Photon,
  \bibinfo{journal}{Science} \textbf{\bibinfo{volume}{355}},
  \bibinfo{pages}{156} (\bibinfo{year}{2017}{\natexlab{a}}).

\bibitem[{\citenamefont{Stockklauser et~al.}(2017)\citenamefont{Stockklauser,
  Scarlino, Koski, Gasparinetti, Andersen, Reichl, Wegscheider, Ihn, Ensslin,
  and Wallraff}}]{Stockklauser2017}
\bibinfo{author}{\bibfnamefont{A.}~\bibnamefont{Stockklauser}},
  \bibinfo{author}{\bibfnamefont{P.}~\bibnamefont{Scarlino}},
  \bibinfo{author}{\bibfnamefont{J.~V.} \bibnamefont{Koski}},
  \bibinfo{author}{\bibfnamefont{S.}~\bibnamefont{Gasparinetti}},
  \bibinfo{author}{\bibfnamefont{C.~K.} \bibnamefont{Andersen}},
  \bibinfo{author}{\bibfnamefont{C.}~\bibnamefont{Reichl}},
  \bibinfo{author}{\bibfnamefont{W.}~\bibnamefont{Wegscheider}},
  \bibinfo{author}{\bibfnamefont{T.}~\bibnamefont{Ihn}},
  \bibinfo{author}{\bibfnamefont{K.}~\bibnamefont{Ensslin}}, \bibnamefont{and}
  \bibinfo{author}{\bibfnamefont{A.}~\bibnamefont{Wallraff}}, Strong Coupling
  Cavity QED with Gate-Defined Double Quantum Dots Enabled by a High Impedance
  Resonator, \bibinfo{journal}{Phys. Rev. X} \textbf{\bibinfo{volume}{7}},
  \bibinfo{pages}{011030} (\bibinfo{year}{2017}).

\bibitem[{\citenamefont{{Bruhat} et~al.}(2016)\citenamefont{{Bruhat},
  {Cubaynes}, {Viennot}, {Dartiailh}, {Desjardins}, {Cottet}, and
  {Kontos}}}]{Bruhat2016}
\bibinfo{author}{\bibfnamefont{L.~E.} \bibnamefont{{Bruhat}}},
  \bibinfo{author}{\bibfnamefont{T.}~\bibnamefont{{Cubaynes}}},
  \bibinfo{author}{\bibfnamefont{J.~J.} \bibnamefont{{Viennot}}},
  \bibinfo{author}{\bibfnamefont{M.~C.} \bibnamefont{{Dartiailh}}},
  \bibinfo{author}{\bibfnamefont{M.~M.} \bibnamefont{{Desjardins}}},
  \bibinfo{author}{\bibfnamefont{A.}~\bibnamefont{{Cottet}}}, \bibnamefont{and}
  \bibinfo{author}{\bibfnamefont{T.}~\bibnamefont{{Kontos}}}, Strong Coupling
  Between an Electron in a Quantum Dot Circuit and a Photon in a Cavity,
  \bibinfo{journal}{arxiv:1612.05214}  (\bibinfo{year}{2016}).

\bibitem[{\citenamefont{van~der Wiel et~al.}(2002)\citenamefont{van~der Wiel,
  De~Franceschi, Elzerman, Fujisawa, Tarucha, and Kouwenhoven}}]{Wiel2002}
\bibinfo{author}{\bibfnamefont{W.~G.} \bibnamefont{van~der Wiel}},
  \bibinfo{author}{\bibfnamefont{S.}~\bibnamefont{De~Franceschi}},
  \bibinfo{author}{\bibfnamefont{J.~M.} \bibnamefont{Elzerman}},
  \bibinfo{author}{\bibfnamefont{T.}~\bibnamefont{Fujisawa}},
  \bibinfo{author}{\bibfnamefont{S.}~\bibnamefont{Tarucha}}, \bibnamefont{and}
  \bibinfo{author}{\bibfnamefont{L.~P.} \bibnamefont{Kouwenhoven}}, Electron
  Transport Through Double Quantum Dots, \bibinfo{journal}{Rev. Mod. Phys.}
  \textbf{\bibinfo{volume}{75}}, \bibinfo{pages}{1} (\bibinfo{year}{2002}).

\bibitem[{\citenamefont{Fujisawa et~al.}(1998)\citenamefont{Fujisawa,
  Oosterkamp, van~der Wiel, Broer, Aguado, Tarucha, and
  Kouwenhoven}}]{Fujisawa1998}
\bibinfo{author}{\bibfnamefont{T.}~\bibnamefont{Fujisawa}},
  \bibinfo{author}{\bibfnamefont{T.~H.} \bibnamefont{Oosterkamp}},
  \bibinfo{author}{\bibfnamefont{W.~G.} \bibnamefont{van~der Wiel}},
  \bibinfo{author}{\bibfnamefont{B.~W.} \bibnamefont{Broer}},
  \bibinfo{author}{\bibfnamefont{R.}~\bibnamefont{Aguado}},
  \bibinfo{author}{\bibfnamefont{S.}~\bibnamefont{Tarucha}}, \bibnamefont{and}
  \bibinfo{author}{\bibfnamefont{L.~P.} \bibnamefont{Kouwenhoven}}, Spontaneous
  Emission Spectrum in Double Quantum Dot Devices, \bibinfo{journal}{Science}
  \textbf{\bibinfo{volume}{282}}, \bibinfo{pages}{932} (\bibinfo{year}{1998}).

\bibitem[{\citenamefont{Liu et~al.}(2015{\natexlab{a}})\citenamefont{Liu,
  Stehlik, Eichler, Gullans, Taylor, and Petta}}]{Liu2015}
\bibinfo{author}{\bibfnamefont{Y.-Y.} \bibnamefont{Liu}},
  \bibinfo{author}{\bibfnamefont{J.}~\bibnamefont{Stehlik}},
  \bibinfo{author}{\bibfnamefont{C.}~\bibnamefont{Eichler}},
  \bibinfo{author}{\bibfnamefont{M.~J.} \bibnamefont{Gullans}},
  \bibinfo{author}{\bibfnamefont{J.~M.} \bibnamefont{Taylor}},
  \bibnamefont{and} \bibinfo{author}{\bibfnamefont{J.~R.} \bibnamefont{Petta}},
  Semiconductor Double Quantum Dot Micromaser, \bibinfo{journal}{Science}
  \textbf{\bibinfo{volume}{347}}, \bibinfo{pages}{285}
  (\bibinfo{year}{2015}{\natexlab{a}}).

\bibitem[{\citenamefont{Stehlik et~al.}(2016)\citenamefont{Stehlik, Liu,
  Eichler, Hartke, Mi, Gullans, Taylor, and Petta}}]{Stehlik2016}
\bibinfo{author}{\bibfnamefont{J.}~\bibnamefont{Stehlik}},
  \bibinfo{author}{\bibfnamefont{Y.-Y.} \bibnamefont{Liu}},
  \bibinfo{author}{\bibfnamefont{C.}~\bibnamefont{Eichler}},
  \bibinfo{author}{\bibfnamefont{T.~R.} \bibnamefont{Hartke}},
  \bibinfo{author}{\bibfnamefont{X.}~\bibnamefont{Mi}},
  \bibinfo{author}{\bibfnamefont{M.~J.} \bibnamefont{Gullans}},
  \bibinfo{author}{\bibfnamefont{J.~M.} \bibnamefont{Taylor}},
  \bibnamefont{and} \bibinfo{author}{\bibfnamefont{J.~R.} \bibnamefont{Petta}},
  Double Quantum Dot Floquet Gain Medium, \bibinfo{journal}{Phys. Rev. X}
  \textbf{\bibinfo{volume}{6}}, \bibinfo{pages}{041027} (\bibinfo{year}{2016}).

\bibitem[{\citenamefont{Liu et~al.}(2014)\citenamefont{Liu, Petersson, Stehlik,
  Taylor, and Petta}}]{Liu2014}
\bibinfo{author}{\bibfnamefont{Y.-Y.} \bibnamefont{Liu}},
  \bibinfo{author}{\bibfnamefont{K.~D.} \bibnamefont{Petersson}},
  \bibinfo{author}{\bibfnamefont{J.}~\bibnamefont{Stehlik}},
  \bibinfo{author}{\bibfnamefont{J.~M.} \bibnamefont{Taylor}},
  \bibnamefont{and} \bibinfo{author}{\bibfnamefont{J.~R.} \bibnamefont{Petta}},
  Photon Emission from a Cavity-Coupled Double Quantum Dot,
  \bibinfo{journal}{Phys. Rev. Lett.} \textbf{\bibinfo{volume}{113}},
  \bibinfo{pages}{036801} (\bibinfo{year}{2014}).

\bibitem[{\citenamefont{Gullans et~al.}(2015)\citenamefont{Gullans, Liu,
  Stehlik, Petta, and Taylor}}]{Gullans2015}
\bibinfo{author}{\bibfnamefont{M.~J.} \bibnamefont{Gullans}},
  \bibinfo{author}{\bibfnamefont{Y.-Y.} \bibnamefont{Liu}},
  \bibinfo{author}{\bibfnamefont{J.}~\bibnamefont{Stehlik}},
  \bibinfo{author}{\bibfnamefont{J.~R.} \bibnamefont{Petta}}, \bibnamefont{and}
  \bibinfo{author}{\bibfnamefont{J.~M.} \bibnamefont{Taylor}}, Phonon-Assisted
  Gain in a Semiconductor Double Quantum Dot Maser, \bibinfo{journal}{Phys.
  Rev. Lett.} \textbf{\bibinfo{volume}{114}}, \bibinfo{pages}{196802}
  (\bibinfo{year}{2015}).

\bibitem[{\citenamefont{Stockklauser et~al.}(2015)\citenamefont{Stockklauser,
  Maisi, Basset, Cujia, Reichl, Wegscheider, Ihn, Wallraff, and
  Ensslin}}]{Stockklauser2015}
\bibinfo{author}{\bibfnamefont{A.}~\bibnamefont{Stockklauser}},
  \bibinfo{author}{\bibfnamefont{V.~F.} \bibnamefont{Maisi}},
  \bibinfo{author}{\bibfnamefont{J.}~\bibnamefont{Basset}},
  \bibinfo{author}{\bibfnamefont{K.}~\bibnamefont{Cujia}},
  \bibinfo{author}{\bibfnamefont{C.}~\bibnamefont{Reichl}},
  \bibinfo{author}{\bibfnamefont{W.}~\bibnamefont{Wegscheider}},
  \bibinfo{author}{\bibfnamefont{T.}~\bibnamefont{Ihn}},
  \bibinfo{author}{\bibfnamefont{A.}~\bibnamefont{Wallraff}}, \bibnamefont{and}
  \bibinfo{author}{\bibfnamefont{K.}~\bibnamefont{Ensslin}}, Microwave Emission
  from Hybridized States in a Semiconductor Charge Qubit,
  \bibinfo{journal}{Phys. Rev. Lett.} \textbf{\bibinfo{volume}{115}},
  \bibinfo{pages}{046802} (\bibinfo{year}{2015}).

\bibitem[{\citenamefont{Childress et~al.}(2004)\citenamefont{Childress,
  S\o{}rensen, and Lukin}}]{Childress2004}
\bibinfo{author}{\bibfnamefont{L.}~\bibnamefont{Childress}},
  \bibinfo{author}{\bibfnamefont{A.~S.} \bibnamefont{S\o{}rensen}},
  \bibnamefont{and} \bibinfo{author}{\bibfnamefont{M.~D.} \bibnamefont{Lukin}},
  Mesoscopic Cavity Quantum Electrodynamics with Quantum Dots,
  \bibinfo{journal}{Phys. Rev. A} \textbf{\bibinfo{volume}{69}},
  \bibinfo{pages}{042302} (\bibinfo{year}{2004}).

\bibitem[{\citenamefont{Jin et~al.}(2011)\citenamefont{Jin, Marthaler, Cole,
  Shnirman, and Sch\"on}}]{Jin2011}
\bibinfo{author}{\bibfnamefont{P.-Q.} \bibnamefont{Jin}},
  \bibinfo{author}{\bibfnamefont{M.}~\bibnamefont{Marthaler}},
  \bibinfo{author}{\bibfnamefont{J.~H.} \bibnamefont{Cole}},
  \bibinfo{author}{\bibfnamefont{A.}~\bibnamefont{Shnirman}}, \bibnamefont{and}
  \bibinfo{author}{\bibfnamefont{G.}~\bibnamefont{Sch\"on}}, Lasing and
  Transport in a Quantum-Dot Resonator Circuit, \bibinfo{journal}{Phys. Rev. B}
  \textbf{\bibinfo{volume}{84}}, \bibinfo{pages}{035322}
  (\bibinfo{year}{2011}).

\bibitem[{\citenamefont{Kulkarni et~al.}(2014)\citenamefont{Kulkarni, Cotlet,
  and T\"ureci}}]{Kulkarni2014}
\bibinfo{author}{\bibfnamefont{M.}~\bibnamefont{Kulkarni}},
  \bibinfo{author}{\bibfnamefont{O.}~\bibnamefont{Cotlet}}, \bibnamefont{and}
  \bibinfo{author}{\bibfnamefont{H.~E.} \bibnamefont{T\"ureci}}, Cavity-Coupled
  Double-Quantum Dot at Finite Bias: Analogy with Lasers and Beyond,
  \bibinfo{journal}{Phys. Rev. B} \textbf{\bibinfo{volume}{90}},
  \bibinfo{pages}{125402} (\bibinfo{year}{2014}).

\bibitem[{\citenamefont{M\"uller and Stace}(2017)}]{Stace2016}
\bibinfo{author}{\bibfnamefont{C.}~\bibnamefont{M\"uller}} \bibnamefont{and}
  \bibinfo{author}{\bibfnamefont{T.~M.} \bibnamefont{Stace}}, Deriving Lindblad
  Master Equations with Keldysh Diagrams: Correlated Gain and Loss in Higher
  Order Perturbation Theory, \bibinfo{journal}{Phys. Rev. A}
  \textbf{\bibinfo{volume}{95}}, \bibinfo{pages}{013847}
  (\bibinfo{year}{2017}).

\bibitem[{\citenamefont{Nadj-Perge et~al.}(2010)\citenamefont{Nadj-Perge,
  Frolov, Bakkers, and Kouwenhoven}}]{Nadj-Perge2010}
\bibinfo{author}{\bibfnamefont{S.}~\bibnamefont{Nadj-Perge}},
  \bibinfo{author}{\bibfnamefont{S.~M.} \bibnamefont{Frolov}},
  \bibinfo{author}{\bibfnamefont{E.~P. A.~M.} \bibnamefont{Bakkers}},
  \bibnamefont{and} \bibinfo{author}{\bibfnamefont{L.~P.}
  \bibnamefont{Kouwenhoven}}, Spin-Orbit Qubit in a Semiconductor Nanowire,
  \bibinfo{journal}{Nature (London)} \textbf{\bibinfo{volume}{468}},
  \bibinfo{pages}{1084} (\bibinfo{year}{2010}).

\bibitem[{\citenamefont{Mi et~al.}(2017{\natexlab{b}})\citenamefont{Mi, Cady,
  Zajac, Stehlik, Edge, and Petta}}]{Mi2016APL}
\bibinfo{author}{\bibfnamefont{X.}~\bibnamefont{Mi}},
  \bibinfo{author}{\bibfnamefont{J.~V.} \bibnamefont{Cady}},
  \bibinfo{author}{\bibfnamefont{D.~M.} \bibnamefont{Zajac}},
  \bibinfo{author}{\bibfnamefont{J.}~\bibnamefont{Stehlik}},
  \bibinfo{author}{\bibfnamefont{L.~F.} \bibnamefont{Edge}}, \bibnamefont{and}
  \bibinfo{author}{\bibfnamefont{J.~R.} \bibnamefont{Petta}}, Circuit Quantum
  Electrodynamics Architecture for Gate-Defined Quantum Dots in Silicon,
  \bibinfo{journal}{Appl. Phys. Lett.} \textbf{\bibinfo{volume}{110}},
  \bibinfo{pages}{043502} (\bibinfo{year}{2017}{\natexlab{b}}).

\bibitem[{SOM()}]{SOM}
\bibinfo{note}{See the Supplemental Material at [url will be inserted by
  publisher] for details of the theoretical model.}

\bibitem[{\citenamefont{Siegman}(1986)}]{Siegman1986}
\bibinfo{author}{\bibfnamefont{S.~E.} \bibnamefont{Siegman}}, \emph{Lasers}
  (\bibinfo{publisher}{University Science Books, Mill Valley, CA},
  \bibinfo{year}{1986}).

\bibitem[{\citenamefont{Liu et~al.}(2015{\natexlab{b}})\citenamefont{Liu,
  Stehlik, Gullans, Taylor, and Petta}}]{LiuPRA2015}
\bibinfo{author}{\bibfnamefont{Y.-Y.} \bibnamefont{Liu}},
  \bibinfo{author}{\bibfnamefont{J.}~\bibnamefont{Stehlik}},
  \bibinfo{author}{\bibfnamefont{M.~J.} \bibnamefont{Gullans}},
  \bibinfo{author}{\bibfnamefont{J.~M.} \bibnamefont{Taylor}},
  \bibnamefont{and} \bibinfo{author}{\bibfnamefont{J.~R.} \bibnamefont{Petta}},
  Injection Locking of a Semiconductor Double-Quantum-Dot Micromaser,
  \bibinfo{journal}{Phys. Rev. A} \textbf{\bibinfo{volume}{92}},
  \bibinfo{pages}{053802} (\bibinfo{year}{2015}{\natexlab{b}}).

\end{thebibliography}

\end{document}